\documentclass[11pt]{article} 
\newcommand{\myscale}{1.05} 
\usepackage{geometry}
\usepackage{natbib}
\bibpunct[, ]{(}{)}{,}{a}{}{,} 
\usepackage{graphicx}
\usepackage{microtype}
\usepackage{verbatim}
\usepackage{hyperref}
\usepackage{hyperref}
\usepackage{paralist}
\hypersetup{pdfencoding=auto,unicode, psdextra, hidelinks} 
\usepackage[full]{textcomp}
\usepackage[scaled=\myscale]{newtxtext}
\usepackage[scaled=\myscale, varg]{newtxmath}
\usepackage{mathtools, bm}
\usepackage{xcolor, xspace}
\usepackage{setspace}
\usepackage{sidecap}
\usepackage{caption}
\usepackage{afterpage}
\usepackage{float}
\usepackage{sectsty}    
\allsectionsfont{\sffamily}
\newcommand{\W}{\,\text{W}}

\newcommand{\pp}[3][]{\frac{\partial^{#1} #2}{\partial #3^{#1}}}

\newcommand{\dd}[3][]{{\dif^{#1} #2 \over \dif #3^{#1}}}

\newcommand{\dif}{\mathrm{d}}
\newcommand{\dg}{\text{\textdegree}\xspace}
\newcommand{\dgc}{\text{\textdegree\,C}\xspace}
\newcommand{\dgn}{\text{\textdegree\,N}\xspace}
\newcommand{\Kv}{\,\text{K}\xspace}
\newcommand{\DDD}[1]{\text{D}{#1} / \text{D} t}
\newcommand{\DD}[1]{{\text{D}#1 \over \text{D} t}}
\newcommand{\cp}{c_{\! p}}
\newcommand{\figref}[1]{Fig.~\ref{#1}}

\newcommand\xb{{\bm{x}}}
\newcommand\ub{\bm {u}}
\newcommand{\fb}{\bm {f}}
\newcommand{\mps}{\m\ps}
\newcommand{\del}{\nabla}

\newcommand{\ps}{\,\text{s}{^{-1}} }
\newcommand{\m}{\,\text{m}}
\newcommand{\km}{\,\text{km}\xspace}
\newcommand{\kg}{\,\text{kg}\xspace}
\newcommand{\eten}[1]{\times 10^{#1}}
\newcommand{\J}{\,\text{J}\xspace}
\newcommand{\s}{\,\text{s}}
\newcommand{\Hc}{\mathcal H}

\newcommand{\qsat}{q_s}

\newlength{\colwidth}
\newcommand{\COT}{CO$_2$}

\title{\bfseries\sffamily The Trouble with Water: \\
Condensation, Circulation and Climate}

\author{Geoffrey K Vallis \\[0.1cm] University of Exeter}
\date{\small Accepted May 25, 2020}

\begin{document}

\maketitle

\vspace*{-8cm} \noindent {\small  In \textit{Eur.\ Phys. J. Plus,}  DOI :10.1140/epjp/s13360-020-00493-7. \\ Focus issue on `\mbox{Fundamental and open issues in climate dynamics'. } }

\vspace*{7cm}

\begin{abstract} 
This article discusses at a fairly basic level a few of the problems that arise in geophysical fluid dynamics and climate that are associated with the presence of moisture in the air,  its condensation and release of latent heat.  Our main focus is Earth's atmosphere but we also discuss how these problems might manifest themselves on other planetary bodies, with particular attention to Titan where methane takes on the role of water. 

Geophysical fluid dynamics has traditionally been concerned with understanding the very basic problems that lie at the foundation of dynamical meteorology and ocean\-ography.  Conventionally, and a little ironically, the subject mainly considers `dry' fluids, meaning it does not concern itself overly much with phase changes.  The subject is often regarded as dry in another way, because it does not consider problems perceived as relevant to the real world, such as clouds or rainfall, which have typically been the province of complicated numerical models. Those models often rely on parameterizations of unresolved processes,  parameterizations that may work very well but that often have a semi-empirical basis. The consequent dichotomy between the foundations and the applications prevents progress being made that has both a secure basis in scientific understanding and a relevance to the  Earth's climate, especially where moisture is concerned. The dichotomy also inhibits progress in understanding the climate of other planets, where observations are insufficient to tune the parameterizations that weather and climate models for Earth rely upon, and a more fundamental approach is called for.  Here we discuss four diverse examples of the problems with moisture: the determination of relative humidity and cloudiness; the transport of water vapor and its possible change under global warming; the moist shallow water equations and the Madden-Julian Oscillation; and the hydrology cycle on other planetary bodies.

\end{abstract}

\setstretch{1.1}

~\\
A sense of falling, like an arrow-shower   \\
Sent out of sight, somewhere becoming rain.  \\
Philip Larkin,  \textit{The Whitsun Weddings.} 

\section{Introduction}

The trouble with water, or at least the one that we shall discuss here, is that it exists in multiple phases. Water vapor in the atmosphere condenses, moving from a gaseous phase to a liquid phase and sometimes to a solid phase. This condensation affects us in a number of ways.  First, the condensation leads to the formation of clouds, perhaps the most interesting phenomena in the sky,  reflecting a good fraction of the Sun's incoming radiation and trapping the outgoing infra-red radiation emitted by the Earth's surface. Second,  condensation leads to rain, the aspect of weather and climate that affects us most (or at least that is talked about most). And third, the condensation releases heat, and that heat affects the circulation of the atmosphere and makes the circulation different from what it might be in a dry atmosphere, especially in the tropics.   The other problem with water, but one that we will not discuss here, is that it is a greenhouse gas, and that greenhouse effect will increase as our planet inevitably warms.  

Our discussion is bound-up with geophysical fluid dynamics, or GFD, the subject that provides the foundation of dynamical meteorology and oce\-an\-ography.\footnote{GFD in its broadest sense refers to the fluid dynamics of all things geophysical, including such things as volcanology, lava flows and mantle convection. In this article we restrict its use to the fluid dynamics of planetary atmospheres and oceans.}  It provides the theoretical basis for understanding the circulation of the atmosphere and ocean of Earth and, more recently, of other planets. Thus,  for example GFD has given us the theory of Rossby waves \citep{Rossby39}  that modulate atmospheric flow on very large scales. It has provided the theory of baroclinic instability \citep{Charney47, Eady49}, which is the instability that drives weather patterns in the atmosphere (with analogs in the ocean), and which is perhaps the fluid instability that most affects the human condition.  it has provided an explanation for the most prominent large scale features of the ocean circulation --- the great ocean gyres,  the Gulf Stream and its Pacific counterpart, the Kuroshio, the thermocline  and the equatorial undercurrent \citep[e.g.,][]{Stommel48,  Welander59, Cromwell53}.  Books have been written called `Geophysical Fluid Dynamics' \citep[e.g.,][]{Pedlosky87} describing these and more phenomena. 

And yet, suppose we ask the apocryphal person in the street what aspect of the weather and climate affects them most, then rain will almost certainly be close to the top of the list of  replies.  Without rain we  would not have the farms that grow food, or the lakes and rivers that bring us freshwater.  We simply could not live in the manner that we do, and most likely we could not live at all.  Why then has the foundational subject (GFD) not embraced  one of the most important of processes? The question is a little disingenuous, for there has been a great deal of work in the past 30 years or so investigating the effects of moist processes on the atmosphere, and we will describe some of that in this article. Still, it is fair to say that moist processes are not yet part of the canon of GFD in the same way that are, for example, quasi-geostrophic theory,  gravity waves or the rotating shallow-water equations. 

This article is meant to provide a perspective on some of the above matters, mainly as regards Earth but also (and for no extra charge) with some discussion ot Titan, where methane takes on the role of water. It is not meant as a review of the effects of water vapour on the climate,  for which see, for example,  \citet{Sherwood_etal10a},  or \citet{Mitchell_Lora16} for Titan. We also do not discuss radiation, microphysics, or smaller scale dynamical phenomena such as hurricanes or moist convection, since discussions on those abound elsewhere. Rather, although this essay does at times review various well-known theories and phenomena, it is meant also to provides a view into what is not known and where both challenges and prospects for progress lie in the years ahead.  
We discuss four discrete but related topics:
\begin{enumerate}
    \item The distribution and variability of relative humidity and the  implications for cloud formation.
    \item Changes in water vapor transport with global warming. 
    \item Moist geophysical fluid dynamics and the Madden--Julian Oscillation
    \item The hydrology cycle of other terrestrial planets, with particular attention to Titan. 
\end{enumerate}    
Topics are deliberately varied and cover a rather wide scope, but the commonality is that  condensation plays an important role in all.  The intent is to show that there really is no escape from thinking about water and other condensibles, no matter where we look in the climate system on Earth and elsewhere.

\section{A Few Basics}

 A particularly important result is the Clausius--Clapeyron (CC) expression for saturation vapor pressure, $e_s$,  namely
\begin{equation}
	\label{cc.1} 
   	\dd {e_s} T =  \frac{L e_s}{R_v  T^2} ,
\end{equation}
where $L$ is the latent heat of vaporization and $R_v$ is the specific gas constant for water vapor. The left-hand side is written as a total derivative because, aside from a negligibly small dependence of $L$ on pressure,  the right-hand side is a function of temperature alone.  If $L$ is constant (in fact it decreases by about 1\% for every 10\dgc rise in temperature)  then \eqref{cc.1} can be integrated to give 
\begin{equation}
\label{cc.2}
    e_s = e_0 \exp\left[\frac{L}{R_v} \left(\frac{1}{T_0} - \frac 1 T \right) \right] 
       \approx e_0 \exp\left[\frac{L}{R_v} \left(\frac{T - T_0}{T_0^2} \right) \right]  = e_0 \exp( \alpha  T')  , 
\end{equation}
 where $\alpha = L/(R_v T_0^2)$ is a constant and $T' = T - T_0$. The exponential increase  is quantitatively accurate  only for small variations of $T$ around $T_0$   but nonetheless the approximation  is a decent one for the range of temperatures commonly encountered in Earth's atmosphere, say from about 240\Kv to 310\Kv, and below those values the water vapor content is in any case very small.  For  $T_0 =280\Kv$ we find $\alpha \approx 0.068 \Kv^{-1}$,  meaning saturation vapor pressure goes up by just under 7\% per degree,  increasing to almost 9\% per degree at $T_0 = 230\Kv$.  In Earth's atmosphere the temperature falls in the vertical by roughly 6\dg/km and consequently  the saturation vapor pressure has an approximate e-folding height of about 2.5 km.

The relative humidity, $\Hc$ is the ratio of the actual vapor pressure to the  saturation vapor pressure.  If the vapor pressure exceeds saturation (that is, if $\Hc  >  1$) then water vapor will, in most atmospheric circumstances, condense and quickly restore the vapor pressure to saturation levels.  Strictly, though, \eqref{cc.1} is derived assuming thermodynamic equilibrium between the vapor and liquid phases of water. In the atmosphere there is no guarantee that such equilibrium holds, in which case values of vapour pressure exceeding the saturation value can occur.  Nevertheless, fast condensation upon saturation is normally a good approximation to make, and is not the main source of problems involving water vapor and the large-scale circulation. 

The specific humidity, $q$, is the ratio of the density of vapor to that of air and it is related to the vapor pressure, $e$,  by
\begin{equation}
	\label{cc.4} 
       q = \frac{\epsilon e}{p-e(1-\epsilon)}  \approx \epsilon {e \over p}.
\end{equation}
where the second expression on the right-hand side holds when $e \ll p$, and  $\epsilon = R_d/R_v \approx 0.62$ is the ratio of the gas constant of dry air to that of moisture.   At constant pressure the saturation value of humidity, $q_s$, increases approximately exponentially with temperature.  The specific humidity is particularly useful because in the absence of condensation or evaporation it is materially conserved; that is $\DDD q = 0$.  Note though, that relative humidity is not conserved, even when a parcel moves at constant temperature. Thus, a parcel can become saturated either by moving to a lower temperature (conserving $q$ but with $e_s$ falling) or by moving to a higher pressure (conserving $q$ and $e_s$ but with $q_s$ falling).

When water does condense, latent heat is released and the temperature of the surrounding air rises according to
\begin{equation}
	\label{cc.3} 
    \cp \Delta T =	L \Delta q,
\end{equation}
where $\cp$ is the heat capacity of air at constant pressure.  
Thus, for a small change in temperature $\Delta T$, saturation vapor pressure and saturation specific humidity change according to
\begin{equation}
	\label{cc.5} 
   	\frac {\Delta q_s } {q_s} = \frac{\Delta e_s}{ e_s} = \alpha \Delta T. 
\end{equation}
At fixed relative humidity changes in $q$ itself obey the same relation. 

It is the fact that saturation values of water vapor content can readily  be reached  in  Earth's atmosphere that distinguishes water vapor from other tracers in the atmosphere.  Any ideal gas has a saturation vapor pressure,  but those for the other main constituents of the Earth's atmosphere are very high at the temperatures that normally occur on Earth. For example, carbon dioxide has a saturation vapour pressure of about 5.7 MPa (57 bars) at 20\dgc, so there is no danger of carbon dioxide rain. On Mars, however,  the low temperatures ($\approx 200\Kv$) lead to a  substantial fraction of the carbon dioxide in the atmosphere condensing every winter.  And on Titan the surface temperature is about $96\Kv$ and methane is a condensible.   The condensation of water vapor (and of methane on Titan) and the concomitant release of heat directly affects the fluid flows in the atmosphere and leads to the difficulties mentioned in the introduction.   To begin, let us first consider water vapor as a passive tracer and the distribution of relative humidity.  

\begin{figure}
    \centering
    \includegraphics[width=0.8\textwidth]{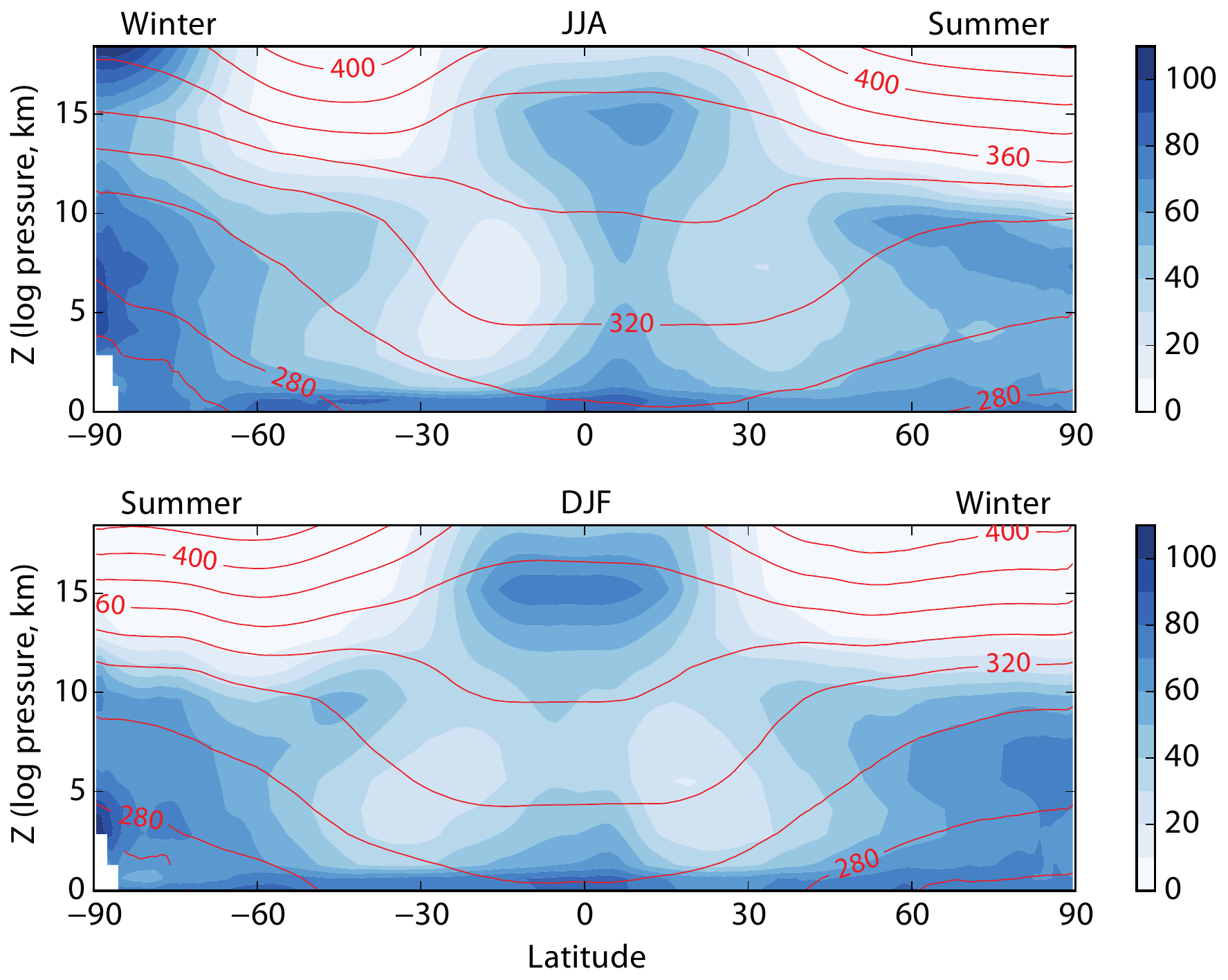}
    \caption{ The zonally-averaged relative humidity (in percent) of Earth's atmosphere for boreal summer and boreal winter (blue shading) and equivalent potential temperature (red contours). }
    \label{fig:rh1}
\end{figure}

\section{Distribution and Variation of Relative Humidity}  \label{sec:distrib}
 
The zonally-averaged relative humidity of the atmosphere is shown in \figref{fig:rh1} and the starting point for an explanation of this pattern begins with the equation for specific humidity, namely
\begin{equation}
	\label{rh.1} 
   	\DD q = E - C.
\end{equation}
Here, $C$ is the condensation and it may be taken as a function that immediately reduces $q$ to its saturated value when $q> \qsat$ and is zero otherwise.   The evaporation, $E$, may be represented as 
\begin{equation}
	\label{rh.2} 
          	E = \kappa \, (q_g - q ) 
\end{equation}
where $q_g$ is the effective humidity of the ground, $q$ is the humidity just above the ground (sometimes taken to the 10 m in atmospheric models) and $\kappa$ is an exchange coefficient, which depends both on the height at which $q$ is taken and the surface wind speed.  Given that $q$ is materially conserved except when evaporation or condensation occurs, the gross features of the relative humidity distribution are recovered by realizing that the relative humidity of a parcel of air is then given by
\begin{equation}
	\label{rh.3} 
   	\Hc(x)  = { q_s\left(T(\xb_\text{sat}), p(\xb_\text{sat}) \right) \over q_s\left(T(\xb), p(\xb) \right) }
\end{equation}
where $q_s\left(T(\xb_\text{sat}), p(\xb_\text{sat})\right)$ is the saturated value of specific humidity at the location where the parcel was last saturated. This formula ignores the effects of re-evaporation of raindrops, which can be large in some circumstances, but  nevertheless, and with a little thought, application of \eqref{rh.3} can be seen to provide an explanation of many of the large-scale features of relative humidity as follows:
 \begin{figure}
     \centering
     \includegraphics[width=0.8\textwidth]{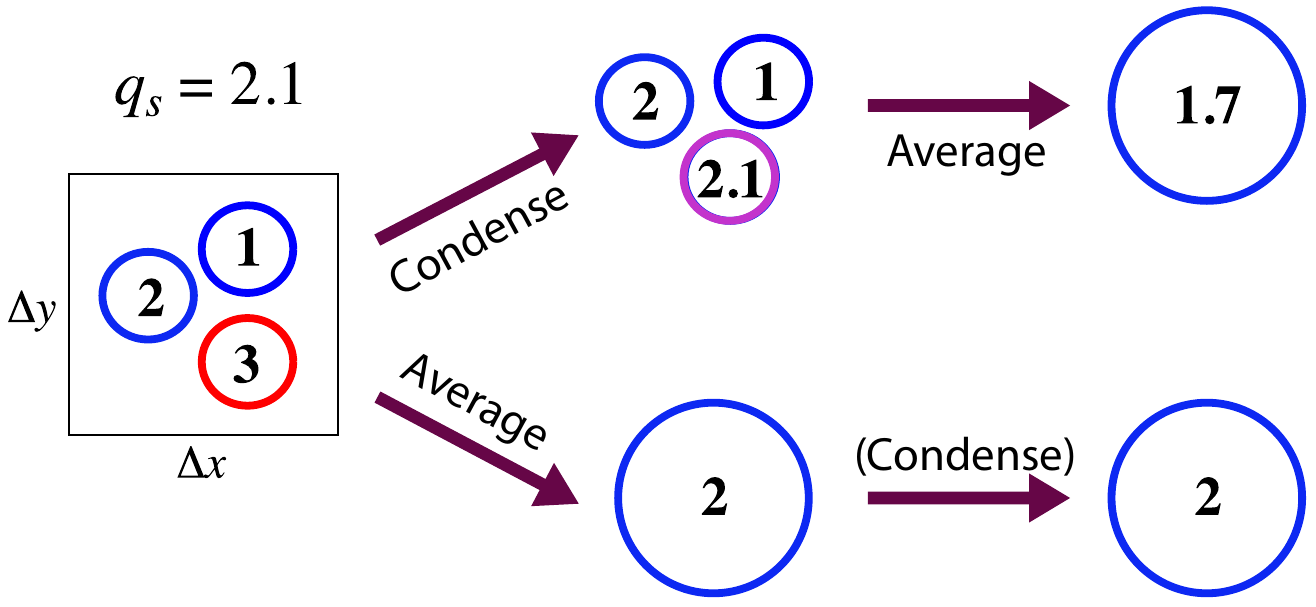}
     \caption{Assume there are three moist parcels, with specific humidities as labelled, in a small volume with local saturation value, $q_s$   In the upper branch we first condense each parcel individually. In the lower branch, we first average over the initial specific humidities  and then carry out the condensation process according to this coarse-grained specific humidity. The  two approaches produce different results, with more moisture  retained when averaging precedes condensation. \citep[Adapted from][]{Tsang_Vallis18}}
     \label{fig:commute}
 \end{figure}   
    \begin{compactitem}
        \item The high levels of relative humidity close to the surface are due to evaporation from the near saturated surface, especially over the ocean and moist ground.
        \item The high levels of relative humidity in the ascending branch of the Hadley Cell arise from upward advection from that nearly saturated surface into cooler air. The branch is not fully saturated on the zonal-average because of the presence of smaller scale motion such as downdrafts that unsaturate the air.
        \item The mid-tropospheric subtropical minimum of relative humidity arises because of the mean descending motion, advecting water vapour into a warmer region and decreasing its relative humidity.
          \item Very low levels of relative humidity in the stratosphere. Since temperature tends to increase above the  tropopause, the tropopause  is a cold trap and relative humidity is very low beyond it. 
        \item The relative humidity tends to increase polewards in mid-latitudes because of the upward and poleward motion of parcels roughly along isentropes  into a cooler environment where it saturates.  Additionally, and not apparent from \figref{fig:rh1},  the  relative humidity varies considerably with longitude in the mid-latitudes because of the chaotic advection by baroclinic eddies.
 \end{compactitem}
  ~

\begin{SCfigure}
     \centering
     \includegraphics[width=0.5\textwidth]{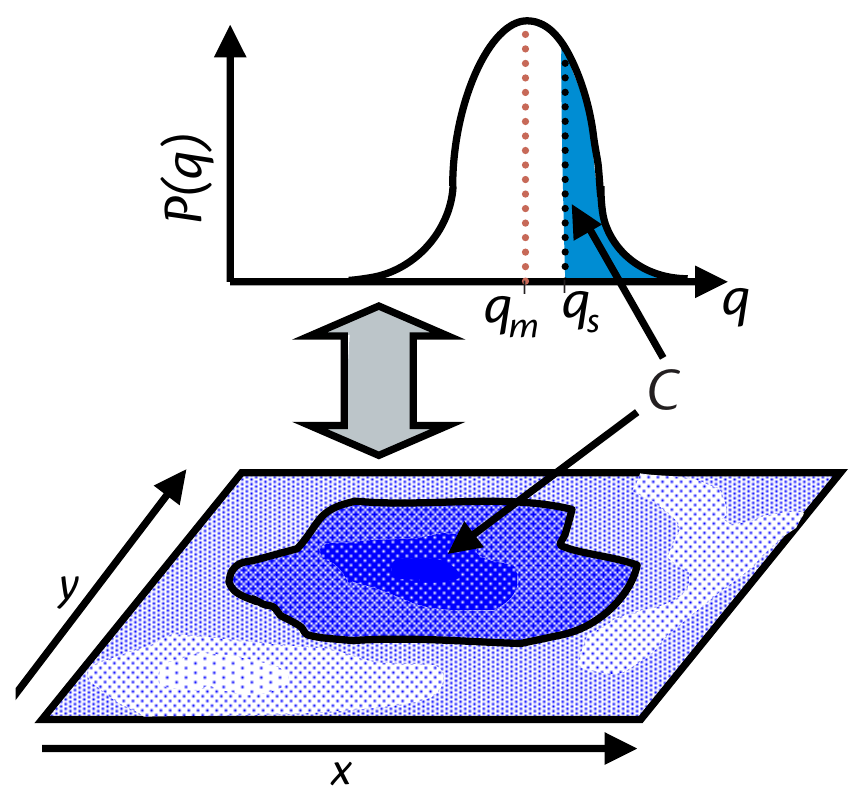}
     \caption{Schematic of a distribution of water content in some volume. The upper plot shows the PDF, with mean value $q_m$ and saturation value $q_s$.  The lower plot is a schematic of the condensation in a grid-box,  with the area integral equal to the amount of saturated water. (Modified from \citet{Tompkins05}.)}
     \label{fig:tompkins}
 \end{SCfigure}

Most modern General Circulation Models (GCMs) can capture  the time-averaged relative humidity reasonably well,  but the distribution (i.e., the PDF) of  water vapour may not be well simulated in global models, with the problems stemming from  finite resolution and the treatment of diffusion, which must be unrealistically large in a GCM.   Figure \ref{fig:commute} illustrates the core of problem, namely that the process of diffusion of water vapor does not commute with the process of condensation.   The immediate consequence is that in a coarse resolution model a given grid box may produce no condensation, because on average the grid box may not be saturated. However, the specific humidity in that grid-box  has a probability distribution, schematically illustrated in \figref{fig:tompkins}, and some of the volume will in fact be saturated and condensation should, in reality, occur.   Consequently,  grid boxes may become saturated too frequently, because diffusion will continuously transport water vapor into regions where  there is a minimum of $q_s$, and this will not be removed efficiently enough by the condensation.

A solution to this problem is to take a probabilistic approaches to the prediction of condensation, an approach that goes back to \citet{Sommeria_Deardorff77} and \citet{Mellor77} using turbulence-closure type arguments. These arguments have been developed and discussed in the years since (see, for example,  \citet{Tompkins02} and \citet{Jakob_Miller02}) and now various more-or-less complicated schemes that use probability distributions of various types are, in fact, in fairly common use in comprehensive GCMs \citep[e.g.,][]{Tiedtke93,Wilson_etal08, Kuwano-Yoshida_etal10}.  The approaches typically make use of models of turbulent diffusion, positing that the PDF has some  particular form and with the parameters of the PDF (e.g., its width and skewness) then determined using a turbulence closure model.  We will now discuss a possible alternative approach, largely following \citet{Tsang_Vallis18}.

 \subsection{A Stochastic Lagrangian model} \label{sec:clouds}

 For simplicity consider a two-dimensional fluid in a box advected by an imposed flow, with streamlines illustrated in panel (a) of \figref{fig:stochastic} and temperature (not shown) diminishing with height.   Moisture enters the system by evaporation from the surface and condenses on saturation. Two kinds of  fluid simulations are illustrated. The left two panels show results from Lagrangian Monte Carlo simulations following individual fluid parcels. That is to say, a large number of parcels are released in a box and are moved by the mean flow, plus an additional stochastic component modeled as a white noise. A parcel is set to be saturated if it collides with the lower boundary but no water enters the domain otherwise.  It then keeps its value of humidity as it wanders through the system, and condensing on saturation.  In figure \ref{fig:stochastic} panel (a) shows a snapshot and panel (b) the relative humidity resulting from a time average of the flow, and for the purposes of this argument they may be regarded as the truth. 
 
 \afterpage{
  \begin{figure}[H]
         \centering
         \includegraphics[width=0.75\textwidth]{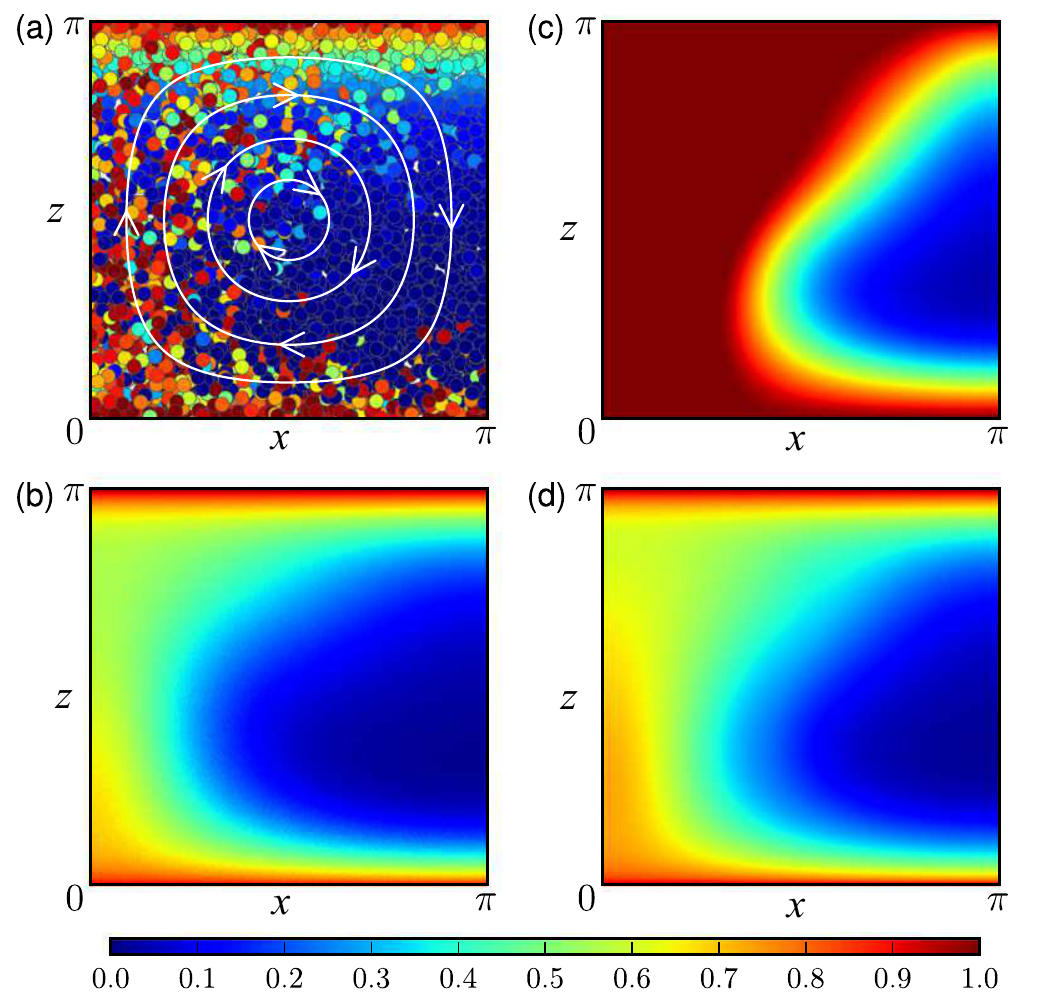}
         \caption{Advection–condensation in an overturning flow. (a) Snapshot of the statistically steady state in a Monte Carlo
         simulation, with color indicating the relative humidity of each parcel. Solid lines are streamlines. (b) Time and bin-averaged relative humidity field  calculated from the simulation in (a). (c) Steady-state relative humidity field r(x, y, t) from a solution   of an Eulerian fluid model with conventional diffusion. (d) Similar to (c) but with a condensation parameterization implemented in the model, as discussed in the text.}
         \label{fig:stochastic}
   \end{figure}   
   }
     
\begin{figure}
     \centering
     \includegraphics[width=0.7\textwidth]{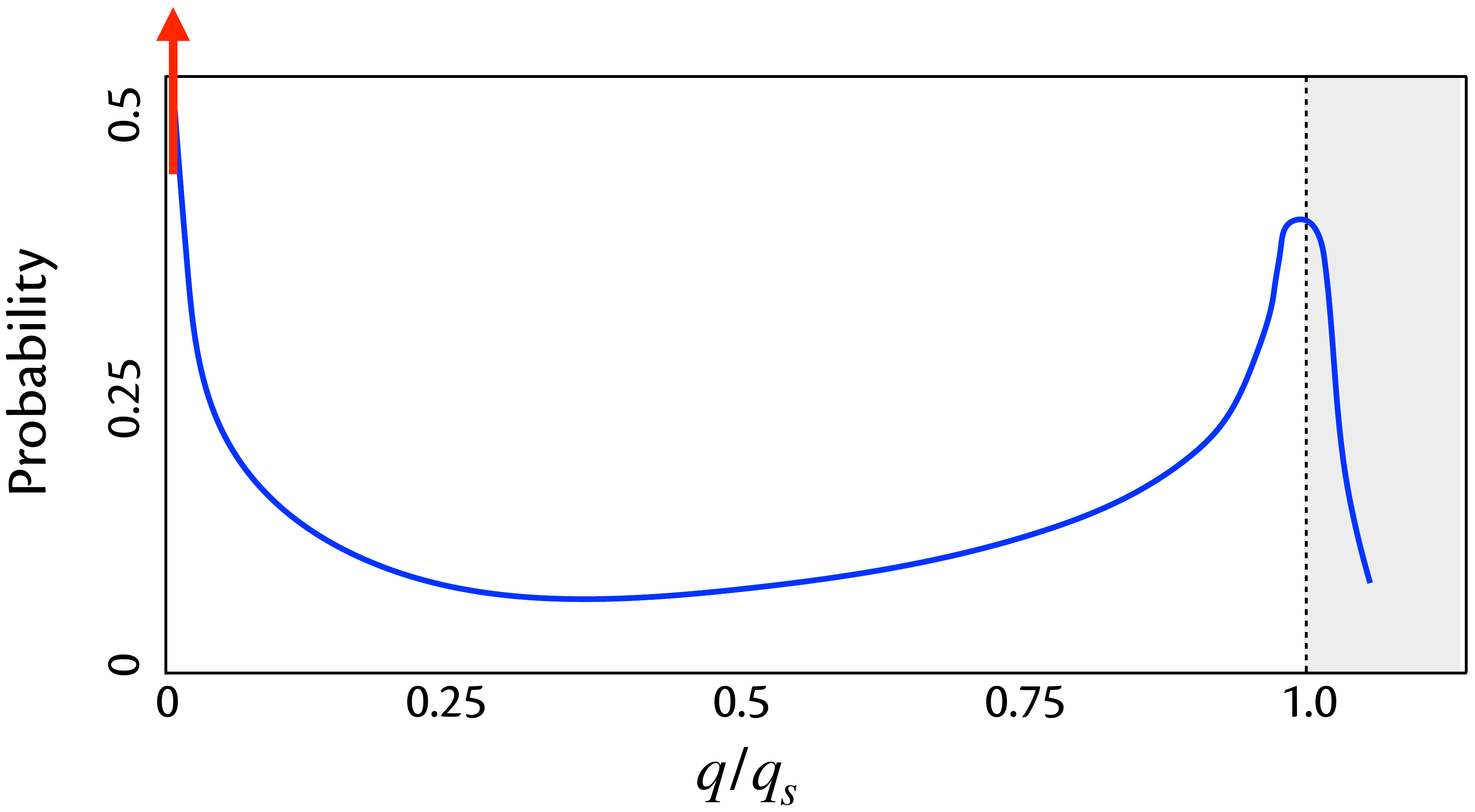}
     \caption{Schematic of a time-averaged probability distribution of relative humidity, $q/\qsat$, at a point.  The  shape and values of the distribution vary,  but the shape is rather generic. There is a dry spike (red arrow) at the minimum value of relative humidity and a broader peak near the saturated value, with a broad minimum at intermediate values.}
     \label{fig:rhpdf}
 \end{figure}

The two right-hand panels of \figref{fig:stochastic} show  the relative humidity calculated from a conventional, Eulerian simulation of the Navier--Stokes equations, with water vapor carried as a passive tracer. Panel (c) uses a conventional diffusivity with a value equivalent  to that produced by the random motion of the parcels in the Monte-Carlo simulation, with condensation upon saturation, and evidently the relative humidity is too high over much of the domain.   (Panel (d) is discussed more below.)   The saturation arises because diffusion will always try to saturate regions that have an interior minimum of $\qsat$ \citep{Pierrehumbert_etal07, Vallis17}.  At the same time, condensation in a coarse Eulerian model  tends to be too inefficient, as discussed above,  and the result is that the model  follows the lower trajectory in \figref{fig:commute} rather than the upper one.   These effects are less apparent at higher resolution and with lower diffusivity, but the resolution of an atmospheric model would have to be extremely high for this issue not to be a problem. 
To obtain a  parameterization for condensation in the Eulerian model that is better  than simple condensation-on-saturation we can use the results of the stochastic model, for this explicitly provides the PDF of relative humidity.   The time-averaged PDF of the generic form illustrated in \figref{fig:rhpdf} is commonly found in observations of the atmosphere \citep{Pierrehumbert_etal07}.  It is typically characterized by having a spike at values of relative humidity near zero, a broader maximum near saturation, and a fairly flat distribution between the two end points. This shape follows from the fact that parcels carry their moisture with them, except when reaching saturation.   Thus, dry parcels (for example those descending from a cold tropopause) will meander through the domain maintaining their dryness, with their relative humidity falling as they enter warmer areas, giving a dry spike near $\Hc = 0$.   On the other hand parcels coming from a warm, moist source (e.g., a warm ocean surface) are nearly saturated, and remain so as they move into colder regions, giving the moist peak to the distribution, with the amplitude of the peak diminishing the further one moves from the source  (see \citealp{Sukhatme_Young11, Tsang_Vanneste17} and \citealp{ Tsang_Vallis18} for more discussion).

\begin{figure}
      \centering
      \includegraphics[width=0.8\textwidth]{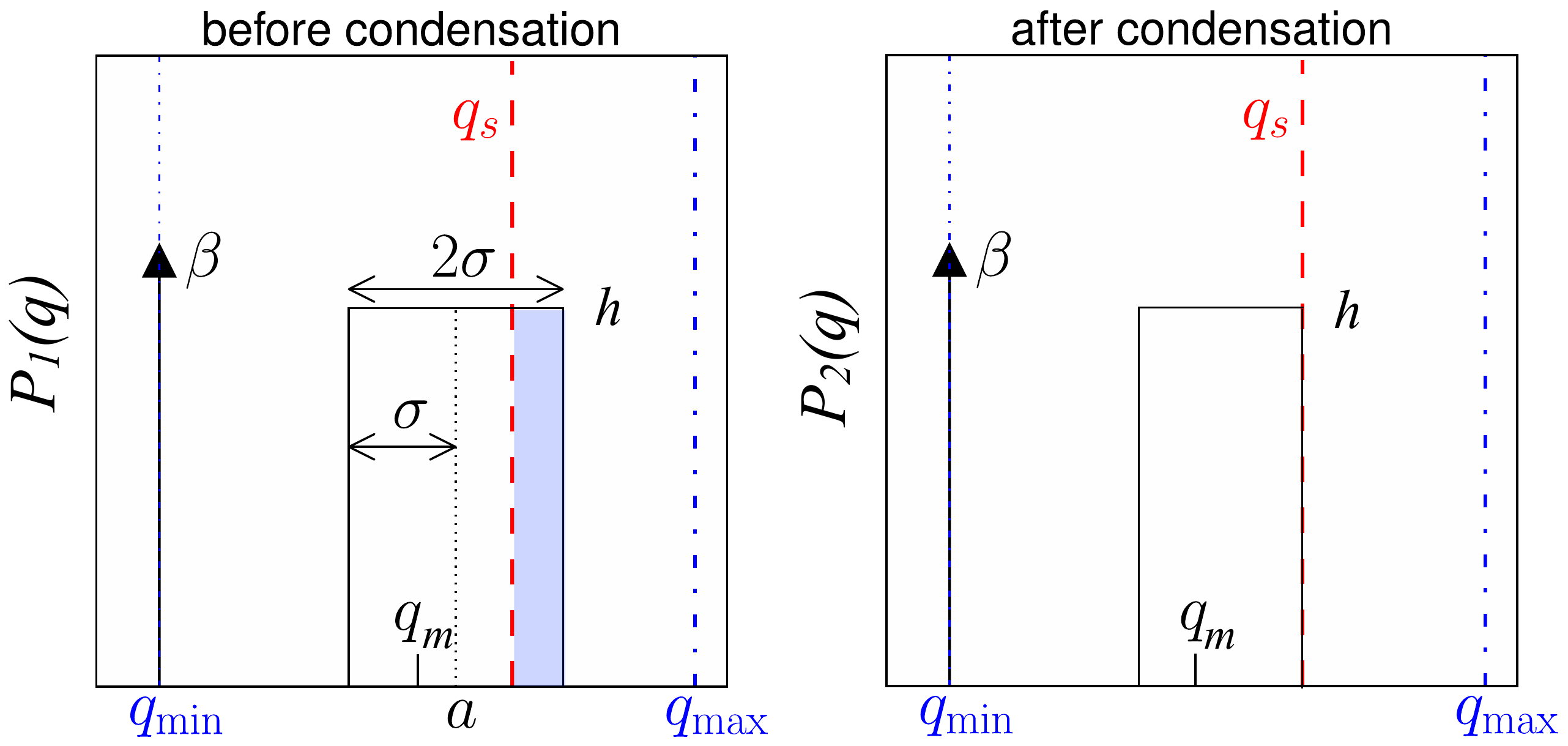}
      \caption{Schematic of the probability distribution of water vapor before and after condensation. The probability distribution is assumed to have a dry spike plus a broad, top-hat shaped distribution of height $h$ and width $2 \sigma$. These parameters are evolved in time, alongside the normal water vapor evolution equation \eqref{rh.1}, which gives the mean of the distribution, with values in the distribution exceed the saturated value being removed each timestep.}
      \label{fig:condschem}
  \end{figure}

If we are able to predict the probability distribution of relative humidity, without actually performing a Monte Carlo integration, then one may use that information to construct a parameterization of moisture condensation in an Eulerian model,  and the relative humidity distribution in \figref{fig:stochastic}d was obtained this way.  Specificallu, we make an ansatz that the probability distribution has a dry spike plus a broad, top-hat shaped distribution at higher humidities, as in \figref{fig:condschem}. The parameters of this distribution (just three in this case) may then be evolved by using them in a Fokker--Planck equation associated with the stochastic Lagrangian model that the individual fluid particles are assumed to obey (i.e., the model illustrated in \figref{fig:stochastic}a).  That is,  to obtain a
prediction of \figref{fig:stochastic}d,   the conventional fluid equations are integrated in time, along with the subsidiary calculation that evolves the statistical moments of the assumed probability distribution.    The fraction of the moisture that is saturated in a given grid box is thereby predicted and that fraction is removed each time step;  condensation then occurs even when the mean value at a given location is not, and the upper pathway in \figref{fig:commute} is then followed.  The prediction of \figref{fig:stochastic}d  is a clear improvement over that of \figref{fig:stochastic}c.

\subsubsection{Future developments}

The stochastic models described above and the more conventional turbulence models have much in common,  and indeed Lagrangian stochastic models are often considered to be the foundation of turbulence dispersion models \citep{Thomson_Wilson13}.  The advantage of the stochastic approach may be that it eliminates the `middle man' -- there is no need to construct a turbulence closure model. Such models necessarily involve the use of ad hoc parameters, and not surprisingly the predictions of condensation and cloud cover (and how it will vary in the future) vary considerably across models. Since  clouds are widely regarded as the greatest single uncertainty in global warming calculations this is an important thing to get right. 

Still, the stochastic Lagrangian methods described above are not without their own set of problems, and extending them to more realistic situations has both fundamental and practical challenges.  Difficulties lie in constructing an appropriate stochastic Lagrangian model appropriate for the full equations of motion, and then in estimating a shape for the PDF that is general enough to cover many eventualities but simple enough to be determined by a small number of parameters.  Those parameters can then be evolved using a truncated Fokker--Plank equation, for it is simply impossible to integrate a full Fokker--Planck equation. One difficulty to note is that  stochastic model must be constructed to take into account the circulation model grid-size, for the time-averaged PDF at a point is not the same as the instantaneous PDF over a grid box, and the smaller the grid the tighter the distribution.

Since this essay is meant to point out future directions, and solving difficult problems is not part of its remit, we will stop this discussion here and move on to a more macroscopic problem,  that of how the water vapor transport, and the circulation that effects that transport, might change as the climate gets warmer.

\section{Changes in Transport of Water Vapor with Warming}  \label{sec:future}

If relative humidity stays roughly constant, then not only will moisture levels increase substantially as temperature rises (with global warming), but so will the magnitude of the moisture transport,  $W$ . This gives rise to the well-known `wet gets wetter' effect \citep{Manabe_Wetherald75, Held_Soden06}.  The argument is a simple one, depending mainly on the assumption that the circulation changes are smaller than the humidity changes, and that relative humidity stays constant.  Begin by noting that if the temperature changes by an amount $\Delta T$,  the specific humidity changes by $\Delta q$ and the humidity transport changes by $\Delta W$, then under CC scaling these quantities are related by
\begin{equation}
	\label{tw.1} 
   	\frac{\Delta W} W = \frac{\Delta q_s} {q_s} = \alpha \Delta T
\end{equation}
where $q_s$ is the saturation value of specific humidity and $\alpha$ was defined in \eqref{cc.2}.   Now, let $\bm W$  be the (vector) vertically integrated transport of humidity such that it is related to precipitation and evaporation by 
\begin{equation}
	\label{tw.2}
   	P - E  = \nabla \cdot \bm W ,
\end{equation}
where $\nabla\cdot$ is the horizontal convergence.  If $W$ satisfies CC scaling then changes in the above quantities are related by 
\begin{subequations}
\begin{align}
	\label{tw.3}
   \Delta (	P - E)  &  = \nabla \cdot ( \Delta \bm W )   \\
                            & \approx \nabla \cdot ( \alpha \Delta T \,  \bm W )   &&  \text {by CC scaling} \\
  &   \approx \alpha \Delta T \, \nabla \cdot \bm W    && \text {if  $\alpha \Delta T$ has small spatial variations}  \\
  & = \alpha \Delta T \, (P - E). 
\end{align}
\end{subequations}
\begin{SCfigure}
    \centering
    \includegraphics[width=0.65\textwidth]{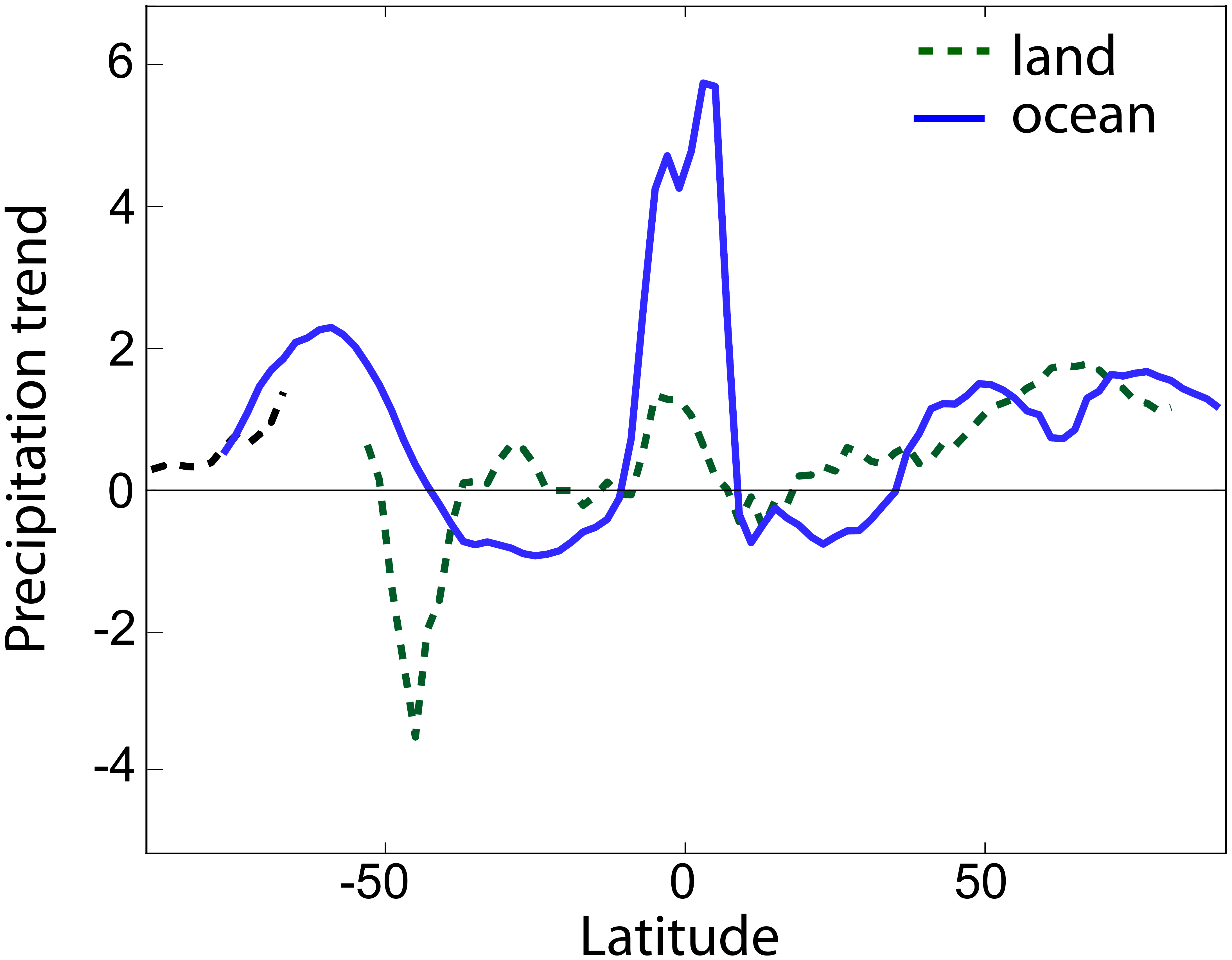} 
    \caption{Trends in precipitation from an ensemble of CMIP5 global-warming simulations with \COT{} concentration increasing by 1\% per year, adapted from \citet{Vallis_etal15}. Units  of the trend are 10\textsuperscript{-2} mm/day per decade.}
    \label{fig:zonalP}
\end{SCfigure}  
Thus, changes in $P - E$ become larger as temperature increases.   The assumptions leading to this result are rather strong ones  -- the relative humidity and the structure of the transport are assumed to have small variations -- and these assumptions do not necessarily hold  over land,  as was noted by \citet{Held_Soden06} and  discussed by \citet{Chou_etal09} \citet{Byrne_OGorman15}, \citet{Pietschnig_etal19} and others.  A basic illustration of the differences over land and ocean is illustrated in Fig. \ref{fig:zonalP} which shows the zonally-averaged change in precipitation over an ensemble of CMIP5 simulations of global warming. Rather clearly one can see that over the ocean the changes roughly correspond to the precipitation pattern itself --- high in the tropics, low in the subtropics --- whereas that pattern is not at all visible over land.

\afterpage{\clearpage
\begin{figure}[H]
    \centering
    \includegraphics[width=0.9\textwidth]{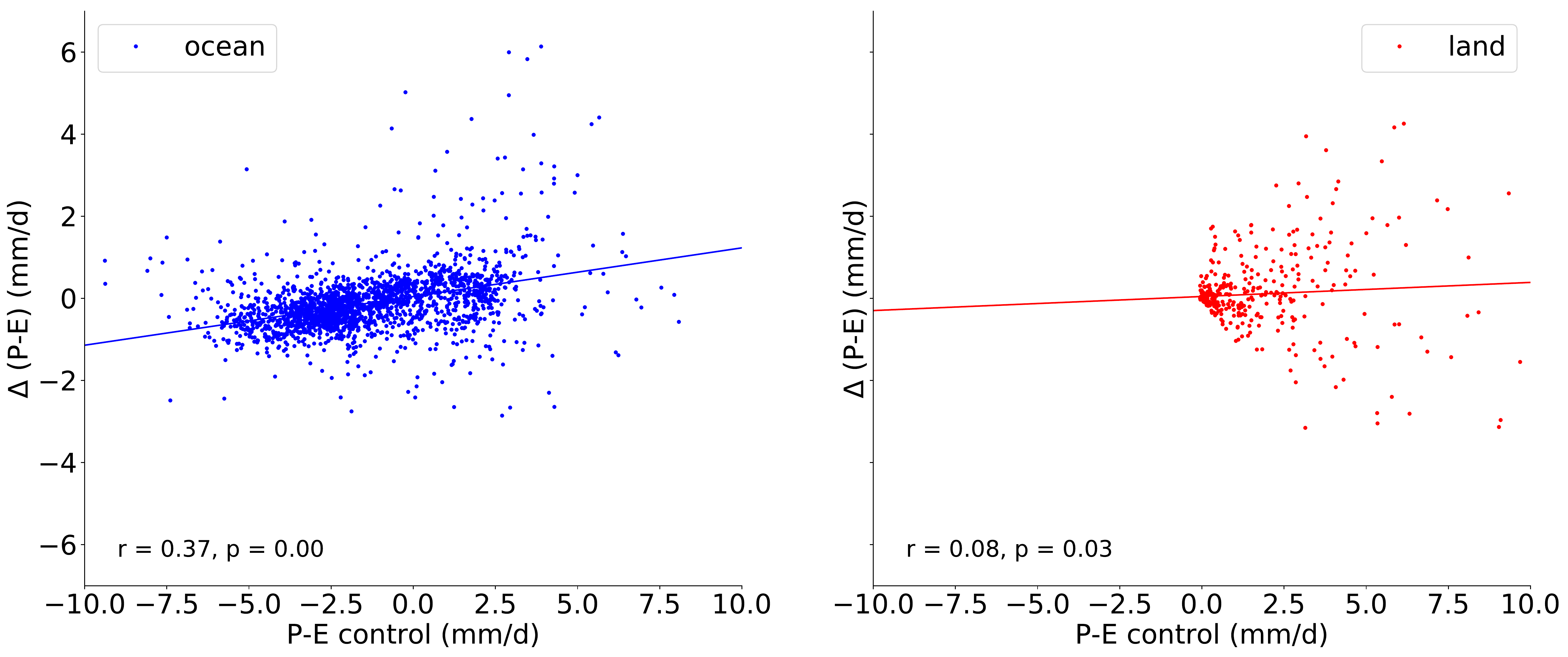} \\
      \includegraphics[width=0.9\textwidth]{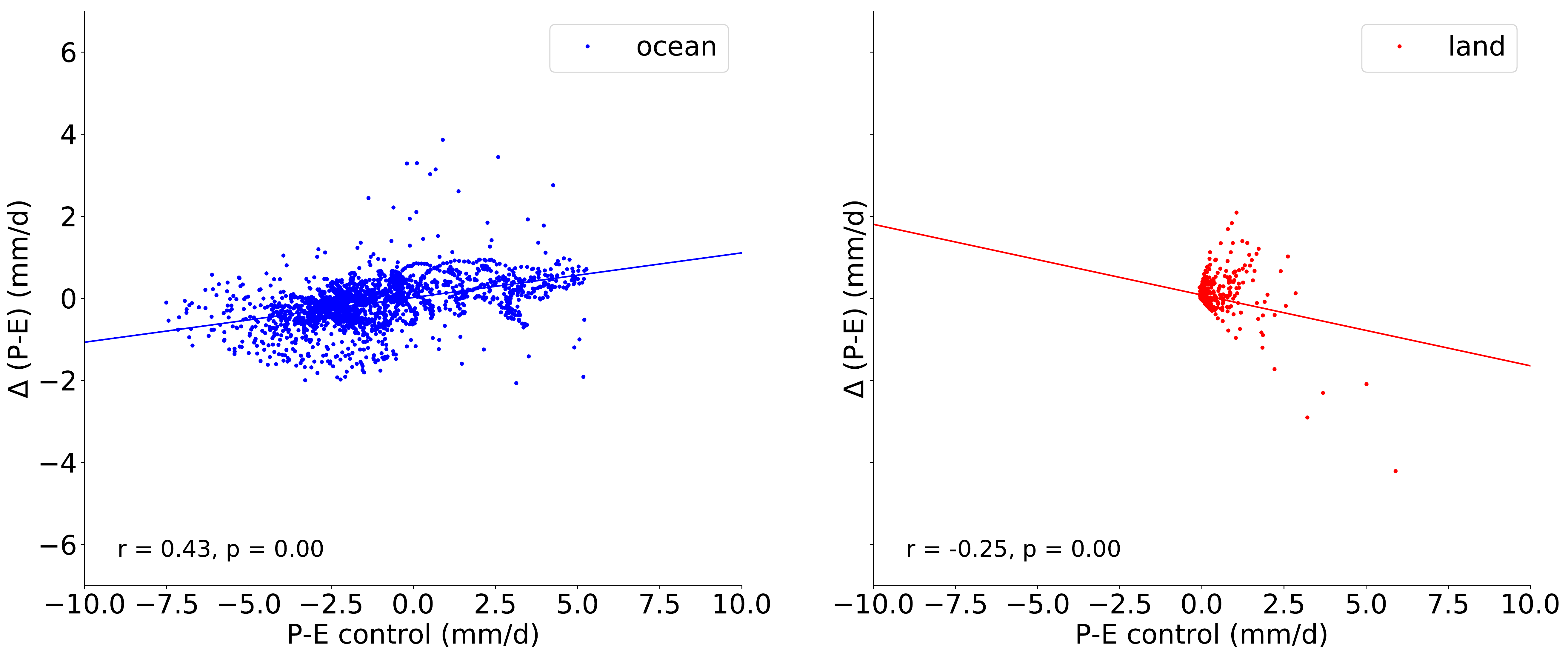} 
    \caption{Scatter plots of $\Delta(P-E)$ vs. $(P-E)$ from a two pairs of  GCM simulations. In each pair one simulation is warmer than the other by about  2.5\dgc on average. Each dot gives the values at a grid point over land or ocean, as labeled. The top  plots comes from simulations with a realistic configuration of land and ocean, whereas the bottom plots come from a very idealized simulation, with continents as in the bottom plot of \figref{fig:P-Emap}.}
    \label{fig:P-Escatter}
\end{figure}    
}

The effect is further  illustrated  in Figures \ref{fig:P-Escatter} and \ref{fig:P-Emap},  which show the  change in $P - E$ in various  GCM simulations taken from \cite{Pietschnig_etal19}, using either a realistic distribution of continents or a very idealized one.   The control simulations have prescribed sea-surface temperatures similar to those observed and in the perturbed simulations the SST is increased uniformly by 2.5\dgc and carbon dioxide doubled from 300 ppm to 600 ppm.  The scatter plots in \figref{fig:P-Escatter} show that, over the ocean, the change in  $P - E$ does follow, on average, $P - E$ itself, albeit with considerable scatter, in both sets of integrations.  Given a  temperature increase of about 2.5\dgc we would expect that $\Delta (P-E) \sim (P-E)/20$, and the results are broadly consistent with that scaling.  Clearly, over land the scaling does not hold, and \figref{fig:P-Emap} illustrates the sensitivity of changes over land to circulation changes.  This plot shows the change in precipitation in warming simulations with either one or two equatorial continents.  Both sets of simulations show some drying over the subtropical ocean where it is already dry, but the change in precipitation (and in $P - E$) over the 'American' continent is very different in the two cases.  The simple reason is that the circulation changes are large, and different in the two cases. In the case with two continents (the lower panel in \figref{fig:P-Emap}) the drying over the smaller, more western continent can be attributed to the  downwelling induced by the enhanced precipitation in the large, eastern continent (described more fully in \citealt{Pietschnig_etal19}). 

\afterpage{
\begin{SCfigure} 
    \centering
    \includegraphics[width=0.6\textwidth]{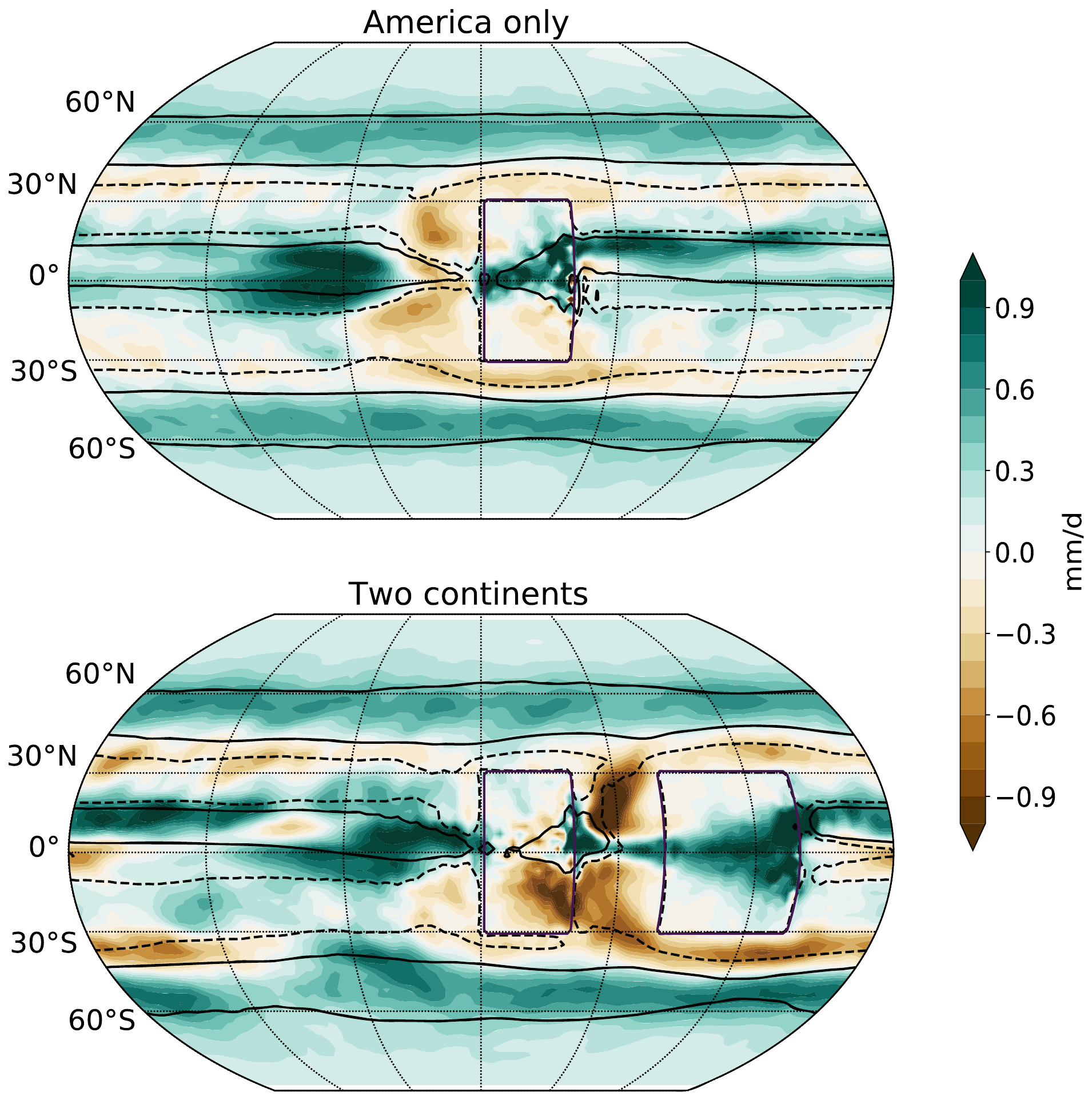}
     \caption{Change in precipitation in two pairs of simulations due to a warming of about 2.5\dgc on average. (The change in $ P- E$ has a similar pattern.)
     In the top plot there is just one  land mass (`America', the rectangular outline) whereas in the bottom plot an additional large continent is added to its west.  The change in tropical circulation and precipitation patterns is very different in two pairs of simulations, and only in the lower one is there drying over the America-like continent. (Adapted from \citealp{Pietschnig_etal19}.) }
    \label{fig:P-Emap}
\end{SCfigure}   
}

The simple point to make is that circulation changes can be large and can overwhelm thermodynamic changes, especially but not only over land. Further, those circulation changes may be induced by the latent heat release itself.   Shifts in circulation, for whatever reason, can be quite extensive; thus, for example, \citet{Scheff_Frierson12} argue that model-predicted declines in subtropical precipitation with global warming are mainly due to shifts in circulation; that is, for dynamical not thermodynamical reasons.   Understanding these changes in the general case is a difficult task, but all is not lost; the circulation changes in the \citet{Pietschnig_etal19} simulation can be understood by appealing to the Matsuno--Gill shallow-water solution \citep{Matsuno66,  Gill80} with a heat source due to latent heat release in precipitation.   This suggests that we might explicitly consider how moisture can be incorporated into the shallow water equations, which leads into to our next topic.

\section{Moist Shallow Water Equations} \label{sec:msw}

The shallow water equations are a mainstay of geophysical fluid dynamics, and the moist shallow water equations make a direct link between GFD and hydrology cycle.   There are various approaches to deriving the equations  \citep[e.g.,][]{Yano_etal95, Rostami_Zeitlin18, Vallis_Penn20} with perhaps the simplest version of the moist shallow water equations having the following form:
\begin{subequations}
\label{msw.1}
\begin{align}
    \DD \ub + \fb \times \ub & = - g \del h ,  \\ 
    \pp h t +  \nabla \cdot (h \ub)  & =  - \gamma C , \\   
    \pp q t + \nabla \cdot (q \ub ) & = E - C.  
\end{align}
\end{subequations}
The notation is standard for the  shallow  water equations with the addition of a moisture equation in which $q$ is a specfic humidity,  $E$ represents evaporation from the surface,  $C$ represents condensation and $\gamma$ is an appropriate latent heat of condensation.  Frictional and dissipative terms have been omitted.   Evaporation may be sensibly parameterized via a bulk aerodynamic formula such as
\begin{equation}
	\label{msw.2} 
          	E = \lambda (q_g - q ) \Hc (q_g -  q),    
\end{equation}
where $\lambda$ is an evaporative parameter (that might be made proportion to velocity), $q_g$ is the surface value of humidity and $\Hc$ is the Heaviside function, ensuring negative evaporation (i.e., dew formation) does not occur.  Condensation may be parameterized supposing that water vapor is quickly removed from the system when the fluid is saturated --- the so-called `fast condensation' process, which may be represented by  
\begin{equation}
	\label{msw.4} 
    C  = \Hc (q - \qsat) { (q - \qsat) \over \tau }  ,  
\end{equation}
where $\qsat$ is the saturation specific humidity and $\tau$ is the timescale of condensation (which is taken to be small), and $\Hc$ is again the Heaviside function.  This condensation is assumed to be a source term in the height equation, represented by $- \gamma C$ in (\ref{msw.1}b).  The saturation value of humidity, $\qsat$, might in the simplest case be taken as a constant or, following \eqref{cc.2}, chosen to vary exponentially with geopotential height as
\begin{equation}
	\label{msw.5} 
    \qsat = q_0 \exp(-\alpha' h), 
\end{equation}
where $\alpha'$ is a constant.  More discussion and values for these parameters are given in the papers referenced above.  

One might well ask, why add moisture to the shallow water equations? The reason is that we now have a reasonably simple system with which to explore moist effects.   The shallow water equations themselves have proven a very fecund source of information about the atmosphere and the moist shallow water equations are a simple  way of explicitly combining water vapor into a dynamical framework.  A number of long-standing problems may then become tractable, and here we discuss a new approach to one, the Madden--Julian Oscillation or MJO.

\subsection{Madden--Julian Oscillation} \label{sec:mjo}

The MJO is a large-scale disturbance, centered near the equator and extending meridionally about 20\textdegree\ either side of the equator, that propagates eastward at a few meters per second \citep{Madden_Julian71, Zhang05, Lau_Waliser12}. Ever since its discovery it has been a source of fascination for observers, modellers and theoreticians alike, and although advances have been made no unambiguous mechanism has been identified that has gained the consensus of the community.  Modelling efforts, using either General Circulation Models with parameterizations of convection or convection-permitting models have had mixed success. That is to say, some groups have successfully obtained propagating disturbances of one form or another \citep[e.g.,][]{Liu_etal09, Arnold_Randall15, Khairoutdinov_Emanuel18} but the conditions determining the production of an MJO are not precisely known. Certainly,  the nature of the convective parameterization and  cloud entrainment parameters affect the production of an MJO in models with parameterized convection \citep[e.g.,][]{Holloway_etal13, Benedict_etal14, Klingaman_Woolnough14}.  At the simpler end of the spectrum, the notion that the MJO is some form of `moisture mode' has been explored by various investigators \citep[e.g.,][]{ Raymond01, Raymond_Fuchs09,  Sobel_Maloney13}.  These theories typically assumed that the atmosphere has a simple vertical structure, dominated by the first baroclinic mode,  and so can be modelled using equations similar to the shallow water equations.  A simple representation of moisture is then added and the equations are horizontally  truncated in some way,  keeping just a few modes of variability.  A large-scale, low-order, interaction between the moisture and the flow then emerges, leading to a propagating instability. 

\begin{SCfigure}
    \includegraphics[width=\colwidth]{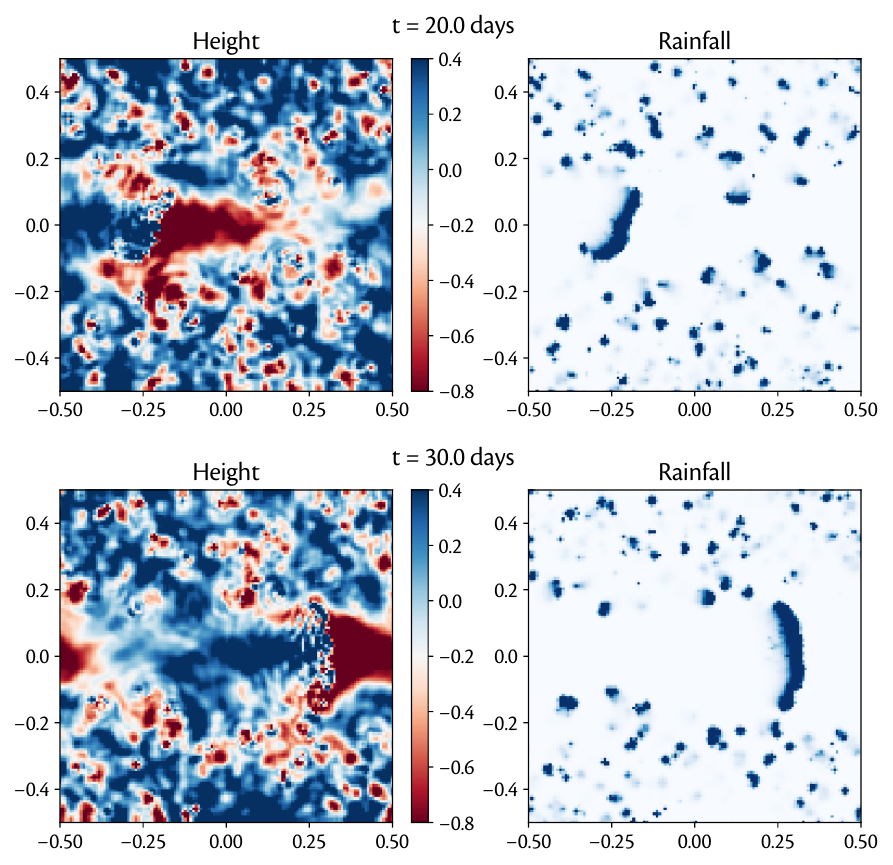} 
    \caption{Snapshots of the height and precipitation fields at the times indicated, in simulations of the moist shallow-water equations on a beta plane. The main disturbance forms about 15 days after initialization and propagates eastward at about 6$\mps$. Units of $x$ and $y$ are $10^7\m$ and the equatorial deformation radius is about $10^6\m$.}
    \label{fig:mjo1}
\end{SCfigure}

\begin{figure}[t]
 \centering
     \includegraphics[width=0.8\textwidth]{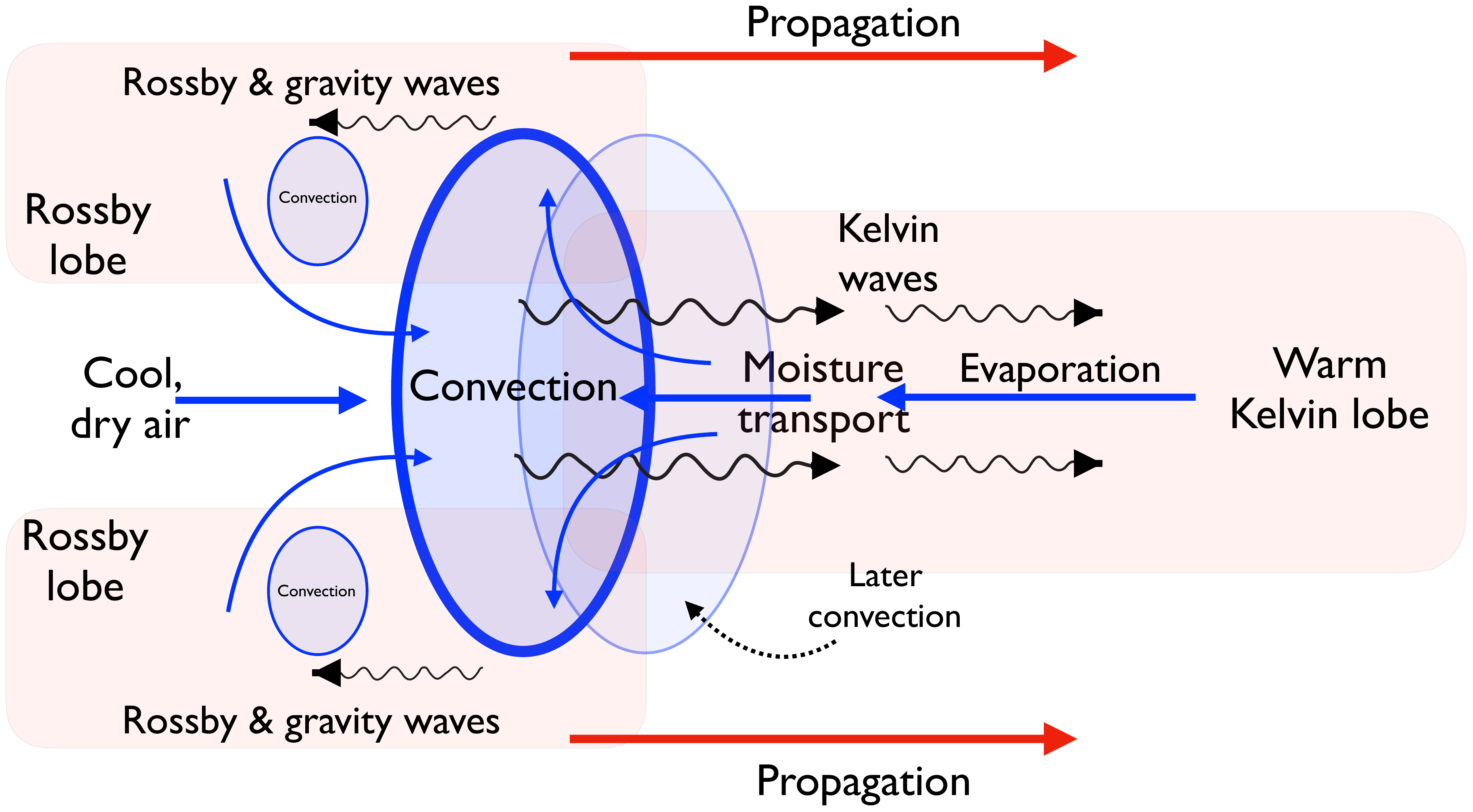} 
     \caption{Schematic of an eastward propagating equatorial disturbance. Convection at the equator gives rise a modified Matsuno--Gill-like pattern. Warm moist air is then drawn in from the east but this is convectively unstable, amenable to triggering by the eastward propagating gravity waves from initial disturbance, and new convection forms on the eastern edge of the original location. The whole pattern then moves unsteadily eastward. (Adapted from \citealt{Vallis_Penn20}.)}
     \label{fig:mjo2}
\end{figure}

A somewhat different approach makes no a priori assumptions about the horizontal scale of the flow and instead seeks to directly solve the moist shallow water equations, \eqref{msw.1}, with the addition of thermal relaxation and dissipative terms.  Simulations give rise to convective organization and eastward propagating structures, as demonstrated in  \figref{fig:mjo1} and  schematized in  \figref{fig:mjo2}. To understand these,  consider first a non-rotating domain and suppose that a perturbation arises that causes the fluid to saturate and condense. This perturbs the height field, giving rise to a converging flow and more condensation, as well as generating gravity waves emitted from the convecting region. These gravity waves may, depending on the degree to which the fluid is conditionally unstable, trigger convection nearby, which in turns triggers more convection and so on.  

Now consider a similar scenario near the equator, with the Coriolis parameter increasing away from the equator as on an equatorial beta-plane.   Condensation at the equator provides a heat source that acts to generate a Matsuno--Gill-like pattern, with Kelvin waves propagating east at the equator and off-equatorial Rossby waves propagating west.  Importantly, such a pattern draws air from the east that, moistened by evaporation from the surface, becomes conditionally unstable.  Convection arises just east of the existing convection and the whole pattern propagates east.    The underlying mechanism is clear: the convection leads to the formation of a large-scale pattern in the geopotential field, which leads to convergence and the  triggering of more convection, creating an excitable system \citep{Meron92, Izhikevich07}.  The equatorial beta-plane gives an east-west asymmetry to the pattern, with longer and warmer `fetch' east of the convection leading to the preferential formation of convection just east of the original site and thence its eastward propagation.  The speed of propagation is related to the rate at which the pattern can progress eastward while maintaining its organization; if the convective center were to move too quickly the convergence of moisture would become  disconnected from the convection itself,  the distinctive pattern would become incoherent and the propagation would cease  \citep{Vallis_Penn20}.   Ultimately, then, the propagation speed is proportional to the speed at which moisture is advected  by the pattern that is set up by the condensation of water; that is, it is an advective velocity rather than modified gravity wave velocity. 

Even though the mechanism above seems robust and straightforward,  more comprehensive models,  in particular General Circulation Models with  convective parameterization schemes,  have not always been able to reproduce it, and one can see why.  If the convection scheme is too sensitive then convection will be excited too easily all around the existing convection and no preferred direction of propagation will emerge.  Similarly, if convection is be triggered too far from the existing site then the coherence cannot then be maintained.  If on the other hand  the convection scheme is too  insensitive then the convective activity may die, perhaps building up again locally in quasi-random locations (because the basic state is convectively unstable) but without creating a coherent pattern.

Other mechanisms of eastward propagation are in principle possible in a similar set-up --- for example,  \citet{Yano_Tribbia17} suggest that the off-equatorial Rossby waves form a non-linear solitary Rossby wave that advects itself eastward.   No matter what the truth is,  the point to be made is that these types of simple model are useful because they can be used to suggest physically plausible mechanisms and that are amenable to exploration by more complex (e.g., convection permitting) models, which in turn may be more directly compared to observations. That is to say, the use of GFD-style ideas, in conjunction with more comprehensive models and observations, gives us a way to make real progress, with solid foundations, on  some complicated real-world problems.  The general area of  the effect of moisture on the large-scale circulation, rather than the somewhat simpler area of the thermodynamic response of moist variables to changes in forcing, is a challenging one. Monsoons  are one area where such effects are almost certainly important, with the latent heat release at monsoon onset potentially initiating a  feedback and enhancing the regime transition that already occurs in dry dynamics, although that remark remains to be tested.  But at least now there are models that include moisture and dynamics in a reasonably simple way and the area seems ripe for progress.

\section{Circulation and Hydrology of Other Planetary Bodies}

Methane is just as troublesome as water, especially at temperatures of 90 K.   Thus,  continuing our theme, we now discuss the  hydrology cycle\footnote
{`Hydrology'  of course derives from the Greek `study of water',  but we  shall extend its meaning to other substances when discussing other planets, and similarly for `moisture'.}   
on other planetary bodies, with particular reference to Titan, where  lakes of liquid methane dot the surface and the weather seems a little British --- weak sunshine, lots of drizzle and the occasional heavy rain-storm.  

The field of study of the atmospheres of other planets is a burgeoning one,  not only of the Solar System planets but increasingly of exoplanets, of which around 4000 are so far known.  It now seems likely that there are billions (indeed billions) of planets in our own galaxy, and of course there are billions of known galaxies, each presumably with their share of planets.  The variety of planets  is enormous, far greater  than that of the stars they orbit, as we see quite clearly in our own Solar System. Planetary atmospheres differ from each other in their mass and composition, their emitting temperature, the size and rotation rate of the planet itself, and in a host of other parameters. In short, planetary atmospheres are enormously complex, and a one-size fits all approach is doomed to failure.  What is to be done?

 \subsection{The Difficulties}

Two main difficulties arise when thinking about planetary atmospheres: their sheer variety, and the sparsity of observations.   The enormous variety suggests that there is no simple theory that can explain them all; there is no equivalent of the Hertzsprung--Russell diagram that  characterizes main-sequence stars.  However, there \textit{are} unifying features and common physical principles that apply across planetary atmospheres, as explored by \citet{Showman_etal10}, \citet{Read11}, \citet{Kaspi_Showman15} and others,  and an approach that treats each planet separately will not reveal their underling commonality.  The sparsity of observations brings another challenge: for Earth we have developed complicated numerical  models with parameterizations that, although based on sound physical principles, are tuned to give realistic answers. Without that tuning the various GCMs would differ considerably from each other and nearly all would differ from Earth itself.  The observations of the other planetary atmospheres  in the Solar System are orders of magnitude more sparse that those for Earth, and the observations of exoplanets are orders or magnitude more sparse still, and tuning the parameterizations is not a viable option. 
Put simply,  there are too many planets to model them one by one and, even if that were not the case, the observations are too sparse  for detailed modeling  and over-fitting the sparse data would be an ever-present danger.   
 
 These considerations require that we take a rather different approach to planetary atmospheres.  We must combine reasoning from basic physical principles with models of appropriate complexity for the  problem at hand,  using case studies from Solar System planets and the best observed exoplanets where possible.  Similar considerations apply  to studies of paleoclimate on Earth, where observations are too sparse and the conditions too different from today's to use highly-tuned GCMs.  
 
 In the rest of this section we will limit ourselves to one aspect of planetary climate, the hydrology cycle, focusing on Titan. In general, a hydrology cycle will exist in a planetary atmosphere if that atmosphere contains a condensible; that is, a gas  that is likely to undergo  phase changes at the typical temperatures within the atmosphere, from the gaseous to the solid or liquid phase.  The hydrology cycle will then be important depending both on the amount of condensible and its latent heat.   Let's see how that works on Titan.

\subsection{Titan and other Terrestrial Planets}

Titan is a moon of Saturn, and in one respect it resembles Earth more than any other planetary body in the Solar System: apart from Earth itself, it is the only  planetary body known to have standing liquid on its surface and, we believe, an active hydrology cycle.   Titan's surface temperature is around 95\Kv, which lies between the triple point of methane, 90.7\Kv, and the boiling point of methane, which is about 117\Kv at Titan's surface where the pressure is about 1.5 bar.  Thus,  any methane will be in liquid form at the surface, and indeed lakes are  observed,  most noticeably at high northern latitudes (at least currently).  This methane evaporates into the atmosphere where it can condense and fall back to the surface, forming a hydrology cycle.   Mars comes close to having a hydrology cycle,  but its surface pressure, only 610 Pa, is generally too low for there to be liquid. The triple point of carbon dioxide is 217\Kv and the surface temperature can be both lower and higher than that, depending on location and season,  ranging from 150\Kv  to 290\Kv.  However,  at pressures less than about 5 bars ($5\eten5$ Pa) carbon dioxide  will sublime (change directly from a gas into a solid) at anything above 195\Kv.  There is considerable water ice on Mars,  most of it in the polar ice caps,  but  very little liquid water at the surface except possibly in small amounts mainly on the north polar ice cap.  There may have been more liquid water in the past,  perhaps 3 or 4 billion years ago, when the atmosphere was (it may be hypothesized) denser and contained more carbon dioxide and/or  sulfur dioxide \citep{Halevy_etal07},  giving both a stronger greenhouse effect and a higher surface pressure,  so allowing liquid water to exist.  Venus also may have had liquid water in the distant past, but went into a runaway greenhouse state as the Sun's output slowly increased (on timescales of hundreds of millions of years) and lost all its water.  

To determine whether the hydrology cycle on Titan is likely to be significant let us put in a few numbers.   For methane,  $ L \approx 5\eten{5}$  J/kg and  $R_v \approx 520 \Kv \kg^{-1}\Kv^{-1}$.  Titan's atmosphere is largely nitrogen for which $\cp \approx 1000 \J\kg^{-1}$ and the  the gravitational acceleration is $g = 1.35\m\s^{-2}$.  The dry adiabatic lapse rate on Titan, $g\cp$ is about  $1.3\Kv \km^{-1}$.
Using \eqref{cc.2}, the approximate Clausius--Clapeyron equation for methane  indicates that the saturation vapour pressure varies approximately as $e_s \approx e_0 \exp(-\alpha_m T')$ where $\alpha_m  = L/(R_v T_0^2) \approx 0.12 \Kv^{-1}$ for $T_0 = 90\Kv$. Thus, saturation methane concentrations in Titan's atmosphere increase by about 10\% per degree Kelvin.  The observed lapse rate is about  $0.7\Kv \km^{-1}$, and therefore atmospheric methane concentrations can be expected to fall by around 10\% per kilometer, giving a scale height of about 10 km.  Since Titan's troposphere extends up to about 40 km most of the methane is in the lower troposphere, just as water vapor is on Earth. 

To understand whether condensation is important we need to know the specific humidity of methane.  At 80\Kv the saturation vapor pressure of methane is about 20 hPa rising to 200 hPa at 95\Kv. The surface pressure of Titan, about 1500 hPa, is  larger than this but only by a factor of 7 or 8,   whereas on Earth the surface pressure due to air (1000 hPa) is 50 or so times the saturation vapour pressure of water at 290 \Kv.    Using \eqref{cc.4} with $\epsilon \approx 0.5$ (for methane and nitrogen),   the specific humidity at saturation of Titan's near-surface atmosphere at 95\Kv is then about 0.07 (70 g/kg), compared to values of 0.02 for Earth's tropics.   This high value means that the effects of condensation may be considerable, even though the latent heat of methane is several times smaller than that of water.   If and when condensation occurs the temperature changes according to $\Delta T = L \Delta q_m/\cp $, and if  $\Delta q = 0.02$ then $\Delta T \sim 10\Kv$,  which is about twice as large as the equator to pole  temperature gradient on Titan.    A similar calculation reveals that the moist adiabatic lapse rate is about $0.6\Kv \km^{-1}$ noticeably smaller than the dry adiabatic lapse rate and similar to that observed, suggesting that moist processes may indeed be an important  determinant of the lapse rate.

Given the above values of mixing ratio, the  total amount of methane on Titan is roughly $2000\kg\m{-2}$, equivalent to about 5 m of liquid methane (which has a density of about $420 \kg\m^{-3}$)  at the surface \citep{Mitchell_Lora16}.   Since the observed lakes cover only a small fraction of  Titan's surface, then unless the lakes are very deep (which is unlikely, given what we know about Titan's geomorphology), or there are very extensive wetlands,  there is far more methane in the atmosphere than on the surface, a very different situation from that for water on Earth. Tropospheric clouds have also been observed \citep{Griffith_etal00,  Roe12}, albeit rather short-lived and mainly in the summer,  varying considerably with season (Titan's year is about 29.5 Earth years),  and only covering a small fraction  of the globe. 

The above elementary  considerations suggest that the methane cycle can be large on Titan, but do not tell us what form it takes. If, for example,  the atmosphere were motionless methane would simply diffuse from the surface into the atmosphere until the atmosphere were saturated,  with the precipitation taking the form of methane drizzle and $\Delta q$ would be smaller than the values used above.  Thus, we may ask:
\begin{enumerate}
    \item Is there in fact an active methane cycle and if so how is it maintained?  More generally, what physical conditions are required to maintain an active hydrology cycle on a terrestrial planet?
    \item How does the hydrology cycle interact with the general circulation? Is the hydrology cycle a mostly passive response to the circulation, or does the hydrology cycle affect the circulation in a significant fashion.    
\end{enumerate}

\subsection{Conditions for a Large-Scale Methane Cycle}

The above questions cannot be answered with thermodynamic reasoning alone but we can make some headway by continuing with some fairly simple arguments.  Note first that an active hydrology cycle can occur even without a large-scale temperature gradient and/or large-scale circulation: if the radiative equilibrium is locally unstable then a radiative-convective equilibrium may result, with local convection and precipitation but no large-scale transport of moisture. There is then no need for any surface transport of fluid, for example by an ocean, since the precipitation and evaporation occur on average in the same place.  The hydrology cycle is then maintained by the excitability of the convection, much as it is in Earth's tropics. 

\begin{figure}[t]
    \centering
    \includegraphics[width=\textwidth]{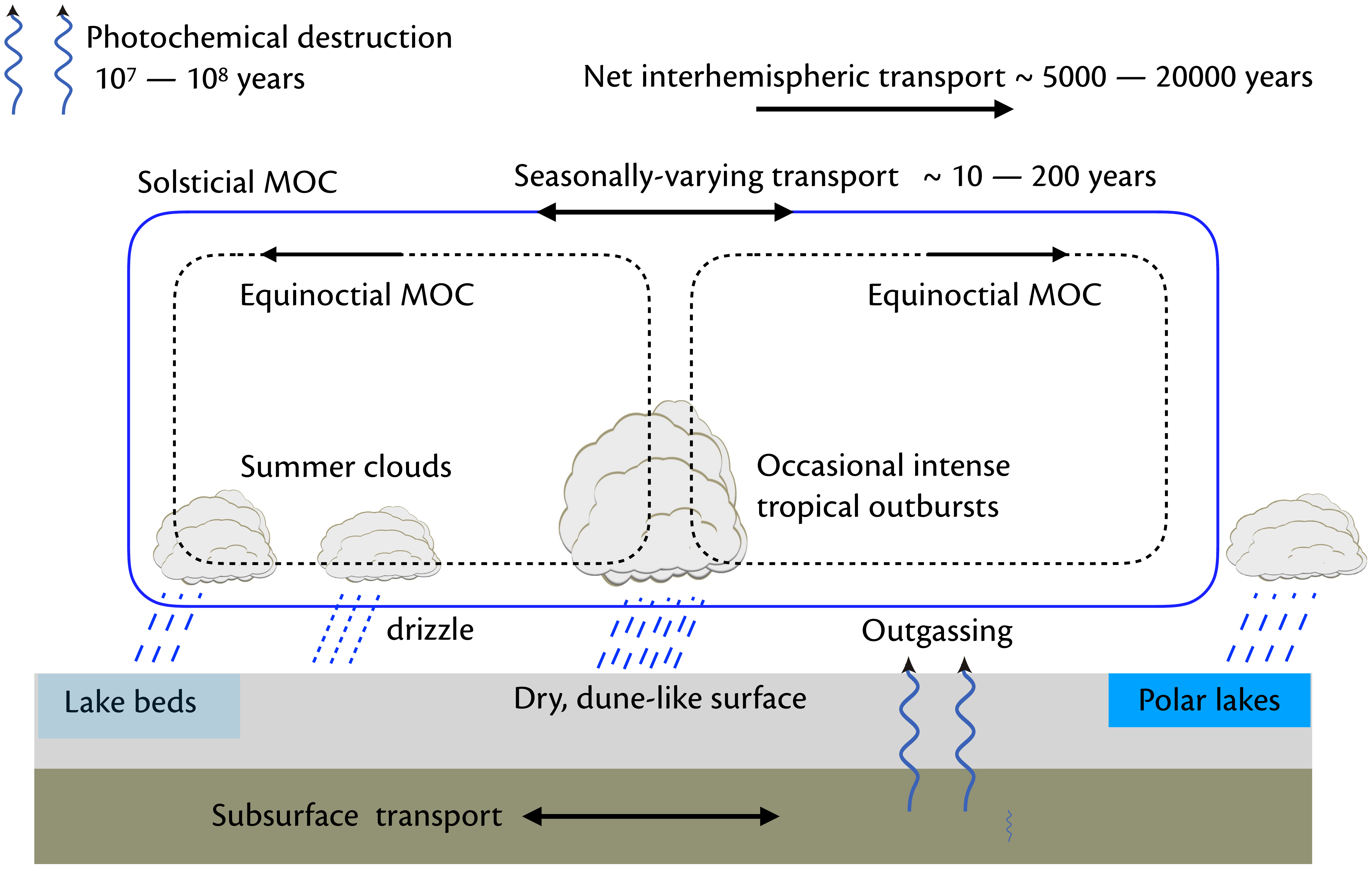}
    \caption{Processes and timescales in Titan's methane cycle. The numbers (in Earth years) refer to how long it will take for the given process to transfer or remove a substantial fraction of the methane content of the atmosphere, and they are uncertain by up to an order of magnitude. The `net interhemispheric transport' refers to the time taken to transport methane from one hemisphere to another because of the hemispheric asymmetry in the solar forcing, which itself varies on about a 45,000 year timescale. The `seasonally-varying transport' refers to the interhemispheric transport over a solsticial season (about 10 Earth years), so only a fraction of a hemispheric inventory is likely depleted in that season before the cycle reverses.}
    \label{fig:titan-methane}
\end{figure}    
 
 A time-averaged energy budget can be a little misleading in this case, since the moist static energy budget then reduces to a balance between top-of-the-atmosphere (TOA) radiative fluxes and surface fluxes of radiation, evaporation and sensible heat.  If the TOA fluxes are small and the surface latent heat and sensible heat fluxes are similar (related by a Bowen ratio) then  single-column radiative-convective calculations \citep{McKay_etal91} give very small evaporative fluxes, but in reality any intermittency in time and space can produce much larger fluxes because of local imbalances. The maintenance of a local hydrology cycle then largely relies on the ability or otherwise of condensation to create gravity waves and trigger condensation nearby in a self-sustaining fashion, as in the excitable systems of Section \ref{sec:msw}.  This in turn depends on the latent heat of condensation and the convective instability of the basic state. 

On Titan, though, there \textit{is} a large-scale circulation and it seems that it is this circulation that drives the methane cycle \citep{Mitchell12, Rodriguez_etal09}. Specifically, there is a large-scale Hadley circulation driven by the pole-to-equator insolation gradient, and because of Titan's obliquity with respect to the Sun, combined with its small radius and low rotation rate (and hence large external Rossby number),  the circulation has a strong interhemispheric component in the solsticial seasons.   The radiative imbalance at the top of the atmosphere varies between about $+0.5\W\m^{-2}$ and $-1\W\m^{-2}$ at high latitudes, implying a polewards flux of energy of up 5 TW. (On Earth, the radiative imbalance is  over 100 times larger than Titan, and the energy flux somewhat less than 1000 times larger,  because Earth is a larger planetary body).   This imbalance drives the meridional overturning circulation, and with it a net transport of methane from low latitudes, where insolation and evaporation are high,  to high latitudes where insolation and evaporation are low \citep{Rannou_etal06,  Schneider_etal12, Lora_etal15}, and hydrology cycle on Titan is most likely driven by this large-scale transport and not by local convective instabilities.   A similar net transport occurs on Earth (albeit with differences because the Hadley Cell has a limited meridional extent) but there the water can be returned by the oceans, and as far as we know there are no oceans on Titan. This leads to a conundrum, as we now discuss.

The methane transport is relatively small compared to the total amount of methane in the atmosphere,  but nonetheless is very significant on longer timescales because Titan's surface is believed to be mostly quite dry, at least at low latitudes.  A surface radiative imbalance of $+0.5\W\m^{-2}$ corresponds, if largely balanced by evaporation,  to a precipitation rate of  about 7 cm/year, which means that any particular atmospheric column (containing 5 or so meters equivalent of liquid methane) would become completely dry on a timescale of a little less than 100 years if not replenished by evaporation. Now, other estimates for precipitation \citep{Barth_Toon06, Tokano_etal06} are  much smaller and if accurate these estimates would extend the drying timescale by up to an order of magnitude. Nevertheless, in any case  the implication is that the methane on Titan should eventually congregate where precipitation exceeds evaporation on timescales of hundreds or possibly thousands of years, which may seem long but by planetary standards is quite short.  On Titan,  we would expect the methane would eventually all aggregate  -- it would be cold-trapped -- at one or both poles. 

If Earth had no ocean basins  similar phenomenon would likely occur there, but on a faster timescale.  A useful rough equivalence for water is that  an  evaporative flux of  $100\W\m^{-2}$ corresponds to about 1 m/year evaporation or precipitation.  Now, an atmospheric column may contain of order  ten  centimeters of liquid water equivalent (far less outside the tropics), so that if the surface were dry the drying time for Earth's atmosphere would be of order a year or less.  The timescale would be much longer if the ground held water, depending on how much water were present: if the groundwater were equivalent to a depth of about 1 meter the timescale for this to be removed by an excess of evaporation over precipitation might be several years or  a few decades,  but nevertheless eventually the hydrology cycle would cease. Unpublished numerical experiments using Isca \citep{Vallis_etal18} configured with land (holding a finite amount of water) over the whole globe, showed that  water first aggregated near the equator on timescales of years, and then on multi-decadal  timescales slowly migrated to and was cold-trapped around the poles.  

These arguments suggest that if a planet has a large scale circulation,  but no ground transport of the liquid condensible, then the hydrology cycle will, eventually but inevitably, die away. Nevertheless, as mentioned there \textit{is} evidence for an active methane cycle on Titan stretching across all latitudes, as sketched in \figref{fig:titan-methane}. Lakes containing liquid methane were observed at high northern latitudes by the Cassini--Huygens mission \citep{Stofan_etal07}, and the \textit{Kraken Mare}, situated around 68\dgn,  may be the largest lake in the Solar System, including those on Earth.  Methane clouds were observed over the South Pole during its summer \citep{Brown_etal02}, and occasional cloud outbursts have been seen in Titan's equatorial region \citep{Schaller_etal09}.  These various observations are consistent with the idea  that it is the global circulation that largely controls the methane cycle and cloud production, possibly with local effects in the tropics driven by convective instabilities.   The difficult question to answer quantitatively is how much ground transport is needed, and if it matters whether the transport is at the surface, for example in wetlands, or below the surface, for example in large aquifers or reservoirs \citep[see][]{Turtle_etal18, Faulk_etal19}.  The strong seasonal cycle on Titan, transporting methane from pole to pole and then back again in the following season,  as well as the possibility of substantial methane storage by lakes,  may reduce the need for ground transport to but a very small fraction of the atmospheric transport in a single season.   However,  this scenario is complicated by an asymmetry between hemispheres:  the  summer solstice is currently close to perihelion, and so slightly warmer but slightly shorter than summer in the northern hemisphere.  The consequence of this is that, over the course of a seasonal cycle, there is a net methane transport to the Northern Hemisphere, possibly accounting for the appearance of lakes in that hemisphere \citep{Aharonson_etal09, Lora_etal14}. However, the accumulation is small and it may take of order 10,000 years to drive most of Titan’s liquid methane to the North Pole. The consequences of this for the need of  ground hydrology are uncertain and it seems that, here at least, the way forward will be through detailed modeling studies and, eventually, in situ observations of Titan's meteorology and geology.  The representation of clouds then becomes important,  bringing us back to the discussion of Section 3, and the need will be for relatively simple,  robust schemes that do not depend heavily on turning, since that is near-impossible for atmospheres other than Earth.  Detailed observations of any sort may never be available for exoplanets, perhaps bringing us a level of irreducible ignorance about planets beyond the Solar System.

\section{Discussion and Future Directions}  

In this essay we've discussed a number of  different topics, ranging from cloud formation to global warming, through the Madden-Julian oscillation to the hydrology cycle on Titan and other planets.  The common thread is that, in all of them,  condensation of water or methane plays an important role.  The point thereby made is that condensation is central to many of the most important topics in climate dynamics.  To conclude I'd like to make some remarks about possible ways forward; these should be regarded as personal opinions rather than a consensus reached by the community. 
\setdefaultleftmargin{5mm}{}{}{}{.5em}{.5em}.
\begin{enumerate}
    \item  It seems important that problems involving phase changes become a more central part of the GFD curriculum, and that `GFD thinking' be applied to such problems.  What is that?  As discussed more in \citet{Vallis16}, GFD is not just a subject area; rather,  is in an approach that,  by tradition, seeks to extract the bare essence of a phenomenon, omitting detail where possible. A GFD approach simply means seeking the most  fundamental explanation of a phenomenon,   sometimes of very complex phenomena, by whatever means possible (including numerical simulations).  In taking such an approach one must eliminate details where possible,  but the question one must then face is `what is a detail?'  In classical GFD phase changes themselves are sometimes  regarded as details or at most character actors, but it is time to promote them to a starring role. We discussed the moist shallow water equations in this article,  but thinking about other idealized moist systems -- moist quasi-geostrophy, the moist Boussinesq and anelastic equations \citep[e.g.,][]{Jusem_Barcilon85, Lapeyre_Held04, Pauluis08, Vallis_etal19} --- as well as the effects of moisture in homogeneous turbulence all deserve more exploration too. 
     Making a better connection between such idealized systems and the messy three-dimensional world will be needed if the idealizations are to be relevant. 
    
    \item Convection is  one area where moisture has played a central role for a long time.  For example, the saturated-adiabatic lapse rate is one of the first things taught in a meteorology class, and extensions such as quasi-equilibrium are staples of tropical meteorology. But understanding the effects of moisture on the large-scale circulation is less well explored. The book by  \citet{Lorenz67} barely mentions moisture, and even today theories of the general circulation do not account for moisture in a systematic way. This is not to say that moisture has been ignored --- \citet{Wang_Barcilon86} and \citet{Fantini03} for example looked at moist effects on baroclinic instability, \citet{Pauluis_etal10} examined how the general circulation looks on moist isentropes, and so on --- but compared to the dry literature it is but a drop.   In particular, the \textit{dynamical} effects of moisture are much more poorly understood than the thermodynamic aspects. The latter, which follow directly from the Clausius--Clapeyron relation, are often quite robust but may be of limited applicability (e.g., may not apply over land).  The former may be dominant but more delicate, with different numerical models giving different results.  Understanding will come from a thorough exploration of simple models as well as relating them to the real world (easily said, of course).
    
    \item The prediction of clouds is a bugbear of General Circulation Models, and indeed almost any numerical model at finite resolution.  Statistical methods, as described in Section \ref{sec:clouds} seem a promising way forward, in part because they avoid the middle-man of turbulence closures. Yet they have two problems: one is that a suitable Lagrangian stochastic model must still be constructed, and the other is that any implementation will be unavoidably complex, perhaps even more so than turbulence models.   Understanding will come if these models can be coupled to more phenomenological recipes, for example by using relative humidity prescriptions with a probability distribution derived, or at least inspired, by using a stochastic model.  Relatively simple models may be all that is warranted for planetary atmospheres other than Earth, in conjunction with the use of convection-resolving models in selected cases, as for example in \citet{Sergeev_etal20}.
    
    \item Many of the above considerations come into play when considering the hydrology cycle on other planets, and we considered Titan as an example. For Titan we have orders of magnitude fewer observations than we do for Earth, and for exoplanets we have orders of magnitude fewer than that.  The construction of complex models for every exoplanet is plainly impossible, and if it were possible it would be almost meaningless. We can proceed by combining basic physical reasoning with models at varying levels of complexity, using case studies and observations where we have them to constrain untrammeled flights of the imagination (although they have their benefits, too).  
    
     Problems abound. For example, it appears -- from the observations that we do have -- that some general circulation models of tidally-locked exoplanets may  under-predict the temperature contrast between day side and night side. This may be because of a cloud cover on the night side, with high clouds having a low emitting temperature, enhancing the contrast in emitting temperature between day and night side. But why should high clouds form on the cold, dark side? Only if we know more about clouds and the hydrology cycle, without using parameterization schemes that are highly tuned to Earth, will we be able to answer that question.   Having a model hierarchy for planetary atmospheres may be still more important for planetary atmospheres than it is for Earth, and the beginnings of one for the Solar System are described in \citet{Vallis_etal18} and \citet{Thomson_Vallis19b}.  
 
\end{enumerate}     
Both for Earth and other planets, understanding will come about through a combination of reductionist and holistic approaches: the former to base processes and models on secure, basic foundations and the latter to understand how the system as a whole works.

\subsubsection*{Acknowledgments}
This work was funded by the Leverhulme Trust,  NERC and the Newton Fund.   This paper draws in part on work of my colleagues and students, and in particular I am grateful to Yue-Kin Tsang, Marianne Pietschnig and James Penn for their help and input.  Many thanks also to Jonathan Mitchell, Spencer Hill, Brett McKim and two anonymous reviewers for their very useful comments.  Any opinions expressed in this article are mine (although they should be yours too). 

 \bibliography{waterrefs}

~\\
\footnotesize \noindent The End.

\end{document}